\documentclass[onecolumn,showpacs,preprintnumbers]{revtex4}
\usepackage{graphicx}
\usepackage{bm}
\usepackage[english]{babel}

\begin{document}

\title{Effect of impurities with retarded interaction
with quasiparticles upon critical temperature of s-wave
superconductor.
}
\author{K.V. Grigorishin}
\email{gkonst@ukr.net} 
\author{B.I. Lev}
\affiliation{Boholyubov Institute for Theoretical Physics of the
National Academy of Sciences of Ukraine, 14-b Metrolohichna str.
Kiev-03680, Ukraine.}
\date{\today}

\begin{abstract}
Generalization of a disordered metal's theory has been proposed
when scattering of quasiparticles by impurities is caused with a
retarded interaction. It was shown that in this case Anderson's
theorem was violated in the sense that embedding of the impurities
in $s$-wave superconductor increases its critical temperature. The
increasing depends on parameters of the metal, impurities and
their concentration. At a specific relation between the parameters
the critical temperature of the dirty superconductor can
essentially exceed critical temperature of pure one up to room
temperature. Thus the impurities catalyze superconductivity in an
originally low-temperature superconductor.
\end{abstract}

 \maketitle

\section{Introduction}\label{intr}

Real superconductors are disordered metal containing various kinds
of impurities and lattice defects. Quasiparticles scatter by these
objects that influences upon superconductive properties of a metal
- critical temperature, gap, critical fields and currents change.
It is well known impurities are two kinds - magnetic and
nonmagnetic. The magnetic scattering differently acts on
components of Cooper pair (for singlet pairing), with the result
that its decay takes place. Superconductive state is unstable
regard to embedding of magnetic impurities - critical temperature
decreases that is accompanied by effect of gapless
superconductivity. In a case of nonmagnetic impurities an ordinary
potential scattering acts on both electrons of a cooper pair
equally. Therefore the pair survives. Hence, \emph{superconductive
state is stable regard to introduction of nonmagnetic impurities -
a gap and critical temperature of a superconductor do not change}.
This statement is Anderson's theorem - $T_{\texttt{C}}$ and
$\Delta(T)$ of an isotropic $s$-wave superconductor do not depend
on presence of nonmagnetic impurities
\citep{degenn,ander,sad,sad1}. This phenomenon is result of the
gap function $\Delta$ and the energy parameter $\varepsilon$ being
renormalized equally \citep{sad}. In a case of anisotropic
$s$-pairing a weak suppression of $T_{\texttt{C}}$ by disorder
takes place \citep{bork,fehren}. For $d$-wave pairing the
nonmagnetic impurities destroy superconductivity like magnetic
impurities \citep{bork,fehren,radt,posazh}. It should be noted
that phonons with lower energies than temperature of the electron
gas are perceived by the electrons as static impurities. Hence the
thermal phonons have no effect on the critical temperature of a
s-wave superconductor that is described by Eliashberg's equations
\cite{ginzb}. However in high-$T_{\texttt{C}}$ oxides the thermal
excitations can break Cooper pairs \cite{gabov} because $d$-wave
pairing takes place. Besides a superconductive state is unstable
regard to introduction of nonmagnetic impurities if the gap is an
odd function of $k-k_{F}$ \citep{sad2}. If pairing of electrons
with nonretarded interaction takes place then $T_{\texttt{C}}$
quickly decreases with an increase of disorder \citep{fay}.

The disorder can influence upon phonon and electron specter in
materials. It results to both increase and decrease of
$T_{\texttt{C}}$. Experiments in superconductive metal showed
suppression of $T_{\texttt{C}}$ by a sufficiently strong disorder
\citep{fiory,hert,bishop,nish}. The strong disorder means that a
free length $l$ is such that $\frac{1}{k_{F}l}\approx 1$ or
$\varepsilon_{F}\tau\approx 1$, where $\tau=l/v_{F}$ is a mean
free time. For weak superconductors as $\texttt{Al}$ or
$\texttt{In}$ a dependence of $T_{\texttt{C}}$ on a disorder
$\frac{1}{k_{F}l}$ has a maximum, but finally the strong disorder
leads to decrease of $T_{\texttt{C}}$ always \citep{belev}, strong
superconductors ($\texttt{Pb}$, $\texttt{Hg}$) have not this
maximum \citep{berg,dyn,belev1}. In the experiments a total
pattern was found: collapse of superconducting state takes place
near Anderson's transition metal-insulator, that is when
$\frac{1}{k_{F}l}\gtrsim 1$. It should be notice that
superconduction appears in amorphus films of $\texttt{Bi,Ga,Be}$
($T_{\texttt{C}}\sim 10\texttt{K}$) just when these materials are
not superconductors in a crystal state \citep{kuz}. In such
systems superconducting is result of intensification of
electron-phonon interaction by disorder. Nowadays universal
mechanisms of influence of a disorder upon $T_{\texttt{C}}$ are
unknown. Several mechanisms of degradation of $T_{\texttt{C}}$
were supposed: a growth of Coulomb pseudopotential $\mu^{\ast}$
\citep{grosu,ander1,bulaev}, influence of the disorder upon a
density of states on Fermi surface $\nu(\xi)$ \citep{sad3,sad4} -
evolution of Altshuler-Aronov singularity \citep{sad,altsh} into
"Coulomb gap". We will not consider these phenomenons  as
violation of Anderson's theorem because they have other nature and
we will consider a weak disorder $\frac{1/l}{k_{F}}\ll 1$ that is
far from a metal-insulator transition.

Introduction of nonmagnetic impurities in a superconductor is
widely used in a practice: the impurities essentially increase a
critical current and a critical magnetic field but do not change
critical temperature at the same time. Our problem is to find such
impurities which violates Anderson's theorem in the direction of
essentiality increasing of the critical temperature
$T_{\texttt{C}}$. Obviously it is matter of nonmagnetic impurities
in a three-dimensional superconductor with s-wave order parameter
$\Delta$. The impurities have to play a role of a catalyst of
superconductivity in an originally low-temperature superconductor.
It should be notice that in an article \citep{zhit} it was shown
that in s-wave superconductors small amounts of nonmagnetic
impurities can increase the transition temperature. However the
correction is of the order of $T_{\texttt{C}}/E_{\texttt{F}}$, and
this effect is result from local variations of the gap function
near impurity sites. Thus the effect is not violation of
Anderson's theorem.

Nowadays a theory of disordered systems has been well developed
for elastic scattering of conduction electrons by impurities
\citep{sad,sad1,pat,altsh}. In a total case the scattering can be
inelastic that is an impurity's potential depends on time
$\upsilon(t)$. In this case to develop a perturbation theory (to
unlink and to sum a diagram series) is impossible. In a section
\ref{retardation} it will be shown that in a special case of
\emph{retarded} interaction with impurities the perturbation
theory can be built and a theory of disordered systems can be
generalized. In a section \ref{Anderson} it will be shown these
impurities violates Anderson's theorem in the direction of
increase of $T_{\texttt{C}}$. A change of the critical temperature
depends on both impurities' parameters and electronic parameters
of a metal matrix. At specific combinations of the parameters the
critical temperature can essentially exceed critical temperature
of a pure metal and has values up to room temperature.

\section{Retarded interaction of conduction electrons with impurities.}\label{retardation}

Let us consider an electron moving in a field created by $N$
scatterers (impurities) which are placed in a random manner with
concentration $\rho=\frac{N}{V}$. A random distribution of the
impurities in a space corresponds to a distribution function:
$P(\textbf{R}_{j})=V^{-N}$. Let a potential of an impurity is a
function of coordinates and time:
$\upsilon(\textbf{r}-\textbf{R}_{j},t)$, where $\textbf{R}_{j}$ is
an impurity's coordinate $\textbf{r}$ is an electron's coordinate.
A total potential created by the impurities is:
\begin{equation}\label{1.1}
    V(\textbf{r},t)=\sum_{j=1}^{N}\upsilon(\textbf{r}-\textbf{R}_{j},t)=
\frac{1}{V}\sum_{\textbf{q}}\sum_{j}\upsilon(\textbf{q},t)e^{i\textbf{q}\left(\textbf{r}-\textbf{R}_{j}\right)},
\end{equation}
where $\upsilon(\textbf{q},t)$ is Fourier transform of the
potential,
$\upsilon(-\textbf{q},t)=\upsilon^{\ast}(\textbf{q},t)$. In most
cases the potential can be considered as point, so that
$\upsilon(\textbf{q})\approx\upsilon=\int\upsilon(\textbf{r})d\textbf{r}$.
Thus the system is spatially inhomogeneous and nonconservative.

\begin{figure}[ht]
\includegraphics[width=10cm]{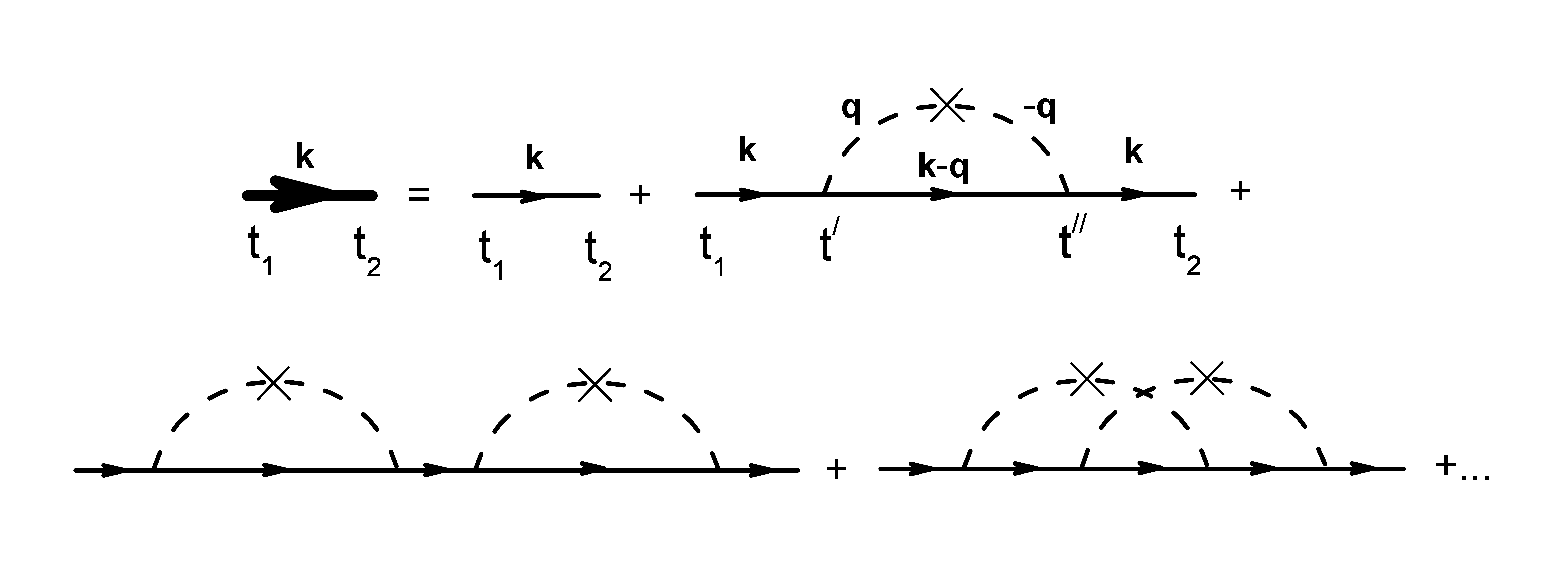}
\caption{The diagram expansion of an averaged Green function
$G(\textbf{k},t)$ in a random field (\ref{1.7}). Dotted lines with
daggers means action of the averaged summarized field of
impurities in a momentum space - a transfer of an intermediate
momentum $\textbf{q}$.} \label{Fig1}
\end{figure}

Considering the potential as weak a perturbation theory can be
constructed writing the secondary quantized interaction
Hamiltonian of an electron with the field (\ref{1.1}) as
$H_{int}=\int
d\textbf{r}\psi^{+}(\textbf{r})V(\textbf{r},t)\psi(\textbf{r})$.
Then a perturbation series for an electron's propagator has a
view:
\begin{eqnarray}\label{1.4}
&&G(1,1')=G_{0}(1,1')+\int d2G_{0}(1,2)V(2)G_{0}(2,1')\nonumber\\
&&+\int d2d3G_{0}(1,2)V(2)G_{0}(2,3)V(3)G_{0}(3,1')+\ldots,
\end{eqnarray}
where $1\equiv(\textbf{r},t),1'\equiv(\textbf{r}',t')$. The
averaging over an ensemble of samples with all possible positions
of impurities recovers spatial homogeneity of a system. In a
representation of secondary quantization the averaging operation
over a disorder has a form \citep{levit}:
\begin{equation}\label{1.6}
G(x,x')=-i\frac{\left\langle\widehat{T}\psi^{+}(x)\psi(x')\widehat{U}\right\rangle_{0}}
{\left\langle\widehat{U}\right\rangle_{0}}\longrightarrow
\left\langle
G(x,x')\right\rangle=-i\left\langle\frac{\left\langle\widehat{T}\psi^{+}(x)\psi(x')\widehat{U}\right\rangle_{0}}
{\left\langle\widehat{U}\right\rangle_{0}}\right\rangle_{\texttt{disorder}},
\end{equation}
where $\widehat{U}$ is an evolution operator,
$\langle\ldots\rangle_{0}$ is done over a ground state of Fermi
system and a lattice (in the numerator and the denominator
separately). The averaging over the disorder is done as follows:
at first the propagator is calculated at the given disorder, and
only then the averaging $\langle\ldots\rangle$ is done (the whole
fraction is averaged). At averaging of the series (\ref{1.4})
$G(\textbf{r},\textbf{r}',t)\rightarrow\left\langle
G(\textbf{r},\textbf{r}',t)\right\rangle$ in a limit
$\rho\rightarrow\infty,\upsilon^{2}\rightarrow
0,\rho\upsilon^{2}=\texttt{const}$ the averages appear with
factorized correlators:
\begin{eqnarray}\label{1.7}
    \left\langle
V(\textbf{r}_{1})V(\textbf{r}_{2})\right\rangle=\rho\upsilon^{2}\delta(\textbf{r}_{1}-\textbf{r}_{2}),\qquad
\left\langle V(1)\right\rangle=0,\qquad\left\langle
V(1)V(2)V(3)\right\rangle=0,\ldots\nonumber\\
\left\langle V(1)V(2)V(3)V(4)\right\rangle=\left\langle
V(1)V(2)\right\rangle\left\langle
V(3)V(4)\right\rangle+\left\langle
V(1)V(4)\right\rangle\left\langle V(2)V(3)\right\rangle+\ldots,
\end{eqnarray}
that corresponds to motion of an electron in Gauss random field
with a white noise correlator. Then an electron's propagator is
determined with a sum of diagrams shown in Fig.\ref{Fig1} (a
diagrammatic techniques of averaging over disorder \citep{sad}).
In an analytic form we have (we use rules of diagrammatic
techniques presented in \citep{matt}):
\begin{eqnarray}\label{1.8}
iG(\textbf{k},t_{1},t_{2})=&&iG_{0}(\textbf{k},t_{2}-t_{1})+\int
dt'\int
dt''iG_{0}(\textbf{k},t'-t_{1})iG_{0}(\textbf{k},t_{2}-t'')\nonumber\\
&&\cdot\rho\int\frac{d^{3}q}{(2\pi)^{3}}(-i)\upsilon(\textbf{q},t')
iG_{0}(\textbf{k}-\textbf{q},t''-t')(-i)\upsilon(-\textbf{q},t'')+\ldots
\end{eqnarray}
Here $G_{0}(\textbf{k},t_{2}-t_{1})$ is a free electron's
propagator depending on a time difference (a pure system is
conservative), $G$ is a dressed electron's propagator. Since
potential of an impurity is a function of a \textit{point of time}
$\upsilon=\upsilon(\textbf{q},t)$, then diagrams of higher orders
cannot be uncoupled, and the series (\ref{1.8}) cannot be summed
(energy is not conserved). The series can be summed partially in
the following cases only. In the first case an impurity's
potential does not depend on time $\upsilon=\upsilon(\textbf{q})$.
It means that an electron scatters elastically by impurities. It
is well described by the disordered system theory
\citep{sad,sad1,pat}. Necessary to us concepts of the theory are
presented in Appendix \ref{A}.

In this article we propose another case when the the series
(\ref{1.8}) can be uncoupled and summed partially. The case is
when an impurity's potential is a function of a time difference
between consecutive scatterings. That is an interaction of
electrons with impurities is retarded (advanced). In the first
approximations a dependence of the scattering potential on a time
difference can be considered as harmonic:
\begin{equation}\label{1.9}
\upsilon(\textbf{q},t')\upsilon(-\textbf{q},t'')=\left\{
\begin{array}{cc}
  \upsilon(\textbf{q})\upsilon(-\textbf{q})e^{-i\omega_{0}(t''-t')}; & \texttt{for} \quad t''>t' \\
  \upsilon(\textbf{q})\upsilon(-\textbf{q})e^{i\omega_{0}(t''-t')}; & \texttt{for} \quad t''<t' \\
\end{array}\right\},
\end{equation}
This means that each impurity is a harmonic oscillator with an
eigenfrequency $\omega_{0}$.  For convenience we can unite the
retarded part and the advanced part in one expression:
\begin{eqnarray}\label{1.11}
\upsilon(\textbf{q},t')\upsilon(-\textbf{q},t'')=|\upsilon(\textbf{q})|^{2}\left[\theta_{t''-t'}e^{-i\omega_{0}(t''-t')}+\theta_{t'-t''}e^{i\omega_{0}(t''-t')}\right]\nonumber
\end{eqnarray}
Let us substitute Fourier transforms of functions $G_{0}$, $G$ and
$\upsilon(\textbf{q},t')\upsilon(-\textbf{q},t'')$ into the series
(\ref{1.8}). Then we obtain:
\begin{eqnarray}\label{1.15}
iG(\textbf{k},\varepsilon)&=&iG_{0}(\textbf{k},\varepsilon)+
\left[iG_{0}(\textbf{k},\varepsilon)\right]^{2}
\rho\int\frac{d^{3}q}{(2\pi)^{3}}\int_{-\infty}^{+\infty}\frac{d\omega}{2\pi}
iG_{0}(\textbf{k}-\textbf{q},\varepsilon-\omega)|\upsilon(\textbf{q})|^{2}(-i)
\frac{2\omega_{0}}{\omega^{2}-\omega_{0}^{2}+2i\delta\omega_{0}}
+\ldots\nonumber\\
&\equiv& iG_{0}(\textbf{k},\varepsilon)+
\left[iG_{0}(\textbf{k},\varepsilon)\right]^{2}
\rho\int\frac{d^{3}q}{(2\pi)^{3}}\int_{-\infty}^{+\infty}\frac{d\omega}{2\pi}
iG_{0}(\textbf{k}-\textbf{q},\varepsilon-\omega)|\upsilon(\textbf{q})|^{2}(-i)
D(\omega)+\ldots
\end{eqnarray}
Thus we can see that \emph{if impurities are harmonic oscillators
with some eigenfrequency $\omega_{0}$ then scattering of electrons
by the impurities is equivalent to scattering of the electrons by
"collective excitations" described with a propagator and a
coupling constant accordingly}
\begin{equation}\label{1.16}
    D(\omega)=\frac{2\omega_{0}}{\omega^{2}-\omega_{0}^{2}+2i\delta\omega_{0}},\qquad \rho|\upsilon(\textbf{q})|^{2}
\end{equation}
Since really in a system such "collective excitations" do not
propagate then we will call the function $D(\omega)$ by
\emph{pseudopropagator}. Thus \textit{in consequence of the
correlations} $\left\langle V(1)V(2)\right\rangle\neq 0$
(\ref{1.7}) \emph{we have a situation as though the "collective
excitations" propagate through a system}. With increasing of
impurities' density $\rho\sim N_{0}/V$ ($N_{0}$ is number of
lattice sites of the matrix) the pseudopropagator does not pass to
a phonon propagator in new lattice (in an obtained alloy) and the
diagrams (Fig.\ref{Fig1}) - to electron-phonon interaction,
because the pseudopropagator and the diagrams with daggers is a
consequence of above-mentioned correlations but phonon propagators
in a solid is a consequence of quasi-elastic interaction between
atoms in a lattice. This means that theory of a disordered metal
does not describe a transition to an alloy with increasing of
impurities' density. In presented article we consider a low
density $\rho\ll\frac{N_{0}}{V}$.

\begin{figure}[ht]
\includegraphics[width=12cm]{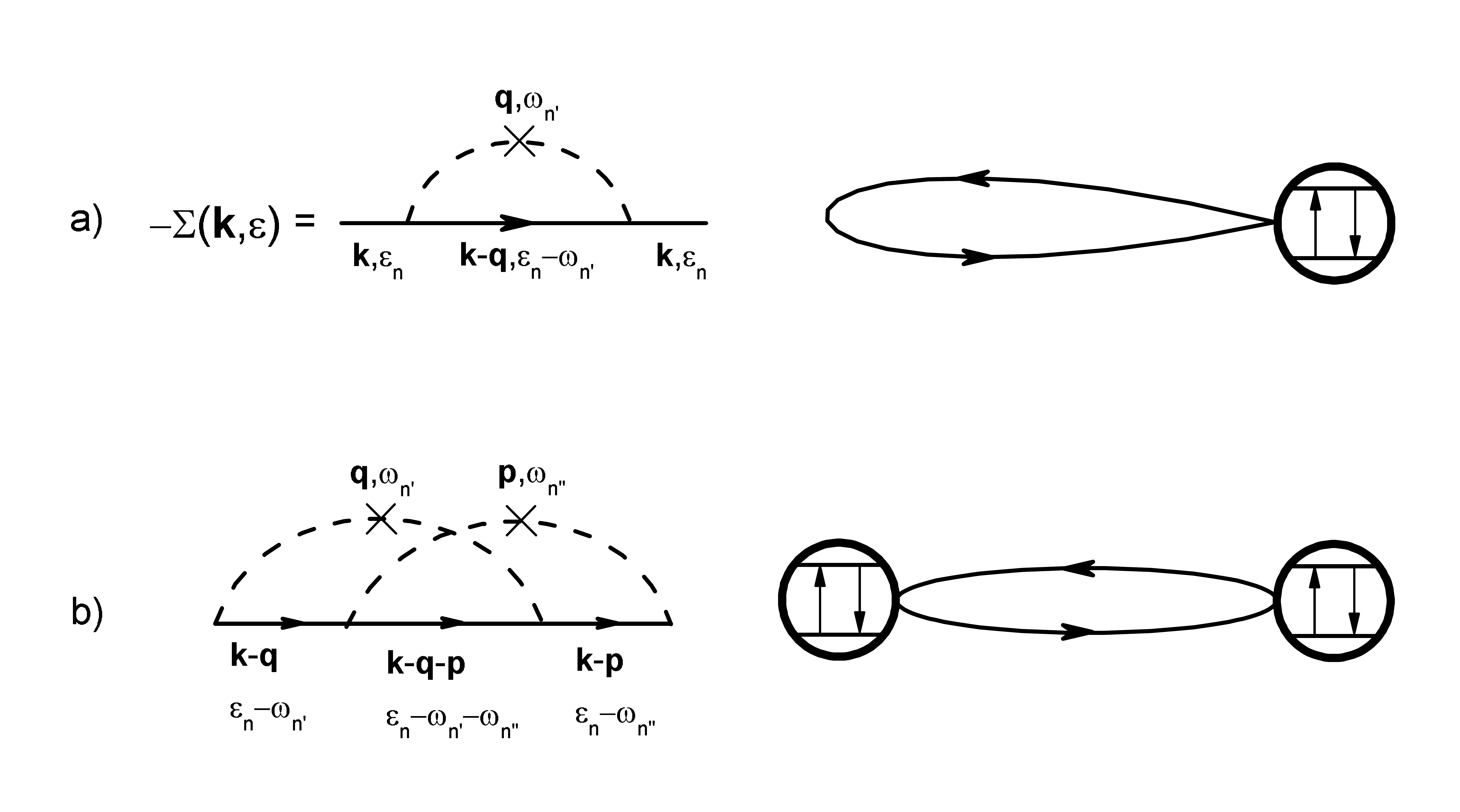}
\caption{Mass operators describing a multiple scattering of
electrons by impurities. These diagrams is analogous to the
diagrams in Fig.\ref{Fig1A}. However now the dotted lines with
daggers mean a scattering by the impurities with energy transfer,
and a multiplier $\rho\upsilon^{2}iD(\omega_{n})$ -
pseudopropagator (\ref{1.16}) is related to them. These processes
can be interpreted with pictures on the right side - the
scattering induces transitions of the impurity-oscillator with
eigenfrequency $\omega_{0}$ between levels.} \label{Fig2}
\end{figure}

High order corrections including any cross processes mentioned in
Appendix \ref{A} is selected similarly in a series (\ref{1.8}). We
can sum the series with standard method and obtain Dyson equation
$\frac{1}{G_{0}}=\frac{1}{G}-i\Sigma$ (in a temperature technics
already). A mass operator is written in a form:
\begin{eqnarray}\label{1.17}
-\Sigma(\textbf{k},\varepsilon_{n})=&&T\sum_{n'=-\infty}^{+\infty}\int\frac{d^{3}q}{(2\pi)^{3}}\rho|\upsilon(\textbf{q})|^{2}
iG_{0}(\textbf{k}-\textbf{q},\varepsilon_{n}-\omega_{n'})iD(\omega_{n'})\nonumber\\
&&+T^{2}\sum_{n'=-\infty}^{+\infty}\sum_{n''=-\infty}^{+\infty}\int\frac{d^{3}q}{(2\pi)^{3}}\int\frac{d^{3}p}{(2\pi)^{3}}
\rho^{2}|\upsilon(\textbf{q})|^{2}|\upsilon(\textbf{p})|^{2}
iG_{0}(\textbf{k}-\textbf{q},\varepsilon_{n}-\omega_{n'})\nonumber\\
&&iG_{0}(\textbf{k}-\textbf{q}-\textbf{p},\varepsilon_{n}-\omega_{n'}-\omega_{n''})
iG_{0}(\textbf{k}-\textbf{p},\varepsilon_{n}-\omega_{n''})
iD(\omega_{n'})iD(\omega_{n''})+\ldots
\end{eqnarray}
Corresponding diagrams are presented in Fig.\ref{Fig2}.
In self-consistent theory the internal electron lines in the
diagrams must be bold $G_{0}\rightarrow G$. Since in metals
$1/k_{F}\simeq a$ ($a$ is a lattice constant) then a weak disorder
corresponds to $\frac{1}{k_{F}l}\ll 1$. A small parameter for the
expansion (\ref{1.17}) is a ratio of a contribution of cross
diagrams to a contribution of diagrams without crossings. In
inelastic scattering by impurities a particle's energy can change
by a value $\triangle\varepsilon\sim\omega_{0}$, that corresponds
to a momentum's uncertainty $\triangle
k=\frac{m\omega_{0}}{k_{F}}$. In \citep{sad} was shown that the
ratio is $\frac{\triangle k}{k_{F}}$, then
\begin{equation}\label{1.19}
\frac{\triangle
k}{k_{F}}=\frac{m\omega_{0}}{k_{F}^{2}}\sim\frac{\omega_{0}}{\varepsilon_{F}}\ll
1.
\end{equation}
This expression likes a situation with phonons where an order of
smallness is an adiabaticity parameter (Migdal's theorem). It is
necessary to notice that the cross diagrams give a small
contribution because, as it was noted in Appendix $\ref{A}$, they
describe an interference contribution in scattering by impurities,
however processes of this type are strongly suppressed in
consequence of an inelastic interaction with the impurities.

\section{Violation of Anderson's theorem.}\label{Anderson}

\subsection{Basic equations.}

Let we have a metal with an attractive interaction between
electrons: $\lambda-\mu^{\ast}>0$ ($\lambda$ is an electron-phonon
coupling constant, $\mu^{\ast}>0$ is Coulomb pseudopotential).
Then the metal can be superconductor. Besides s-wave pairing takes
a place (with zero orbital moment of a pair and zero summary
spin). In the simplest case a superconductive gap is described
with a self consistent equation (it is analogously for $\Delta$):
\begin{equation}\label{2.1}
\Delta^{+}(\varepsilon_{n})=gT\sum_{n'=-\infty}^{+\infty}\int_{-\infty}^{+\infty}
d\xi(-i)F^{+}(\varepsilon_{n'},\xi)w_{\omega_{\texttt{D}}}(\varepsilon_{n},\varepsilon_{n'})
\end{equation}
where $g\approx\lambda-\mu^{\ast}$ is an electron-electron
coupling constant, $F(\varepsilon_{n},\xi)$ is an anomalous
propagator (propagator of a pair):
\begin{equation}\label{2.2}
    F(\varepsilon_{n},\xi)=\frac{i\Delta(\varepsilon_{n})}{(i\varepsilon_{n})^{2}-E^{2}},\quad
    F^{+}(\varepsilon_{n},\xi)=\frac{-i\Delta^{+}(\varepsilon_{n})}{(i\varepsilon_{n})^{2}-E^{2}},
\end{equation}
where $E^{2}=\xi^{2}+|\Delta(\varepsilon_{n})|^{2}$. A function
$w_{\omega_{\texttt{D}}}(\varepsilon_{n},\varepsilon_{n'})$ cuts
the interaction because the pairing interaction is effective if
energies of interacting quasiparticles are less than a
characteristic frequency: $|\varepsilon_{n}|\leq\omega_{D}$. We
can suppose the gap to be real $\Delta=\Delta^{+}$ and to depend
on energy as follows \citep{levit}:
\begin{equation}\label{2.2a}
    \Delta(\varepsilon_{n})=\Delta\frac{\omega_{D}}{\sqrt{\varepsilon_{n}^{2}+\omega_{D}^{2}}}
    \equiv\Delta w_{\omega_{\texttt{D}}}(\varepsilon_{n}).
\end{equation}

\begin{figure}[ht]
\includegraphics[width=12.0cm]{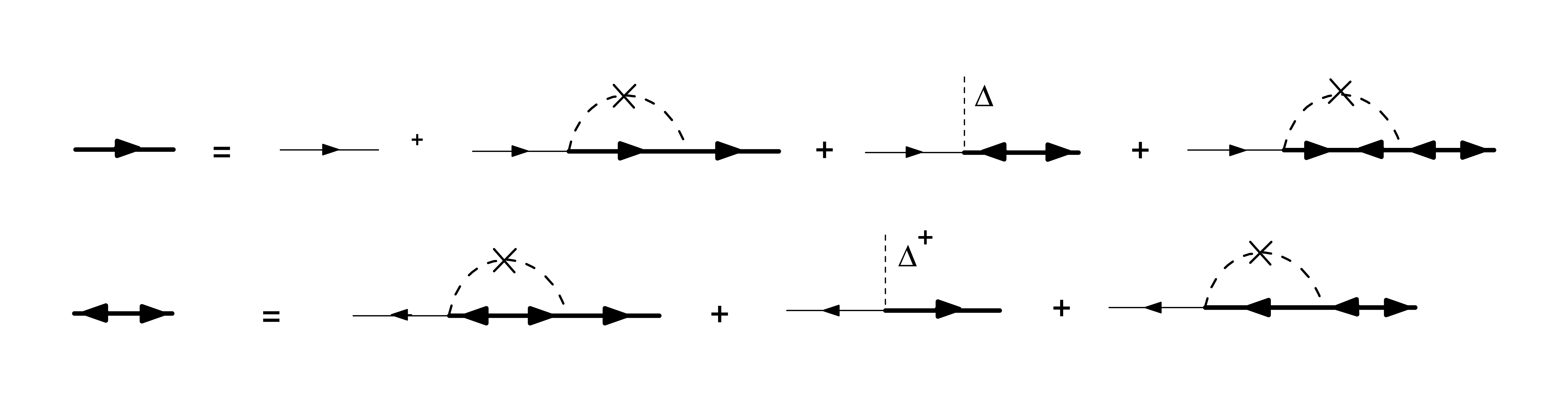}
\caption{Gor'kov equations for a dirty superconductor. Interaction
of quasiparticles with impurities is retarded. A pseudopropagator
of excitations of the impurities $\rho\upsilon^{2}iD(\omega_{n})$
is associated with lines of scattering. Unidirectional thin lines
correspond to free propagators $G_{0}$ (\ref{A.1}) and $G_{0}^{+}$
(with a reverse arrow). Unidirectional bold lines correspond to
dressed normal propagators $\widetilde{G}$ and $\widetilde{G}^{+}$
(\ref{2.4}). Bold lines with oppositely directed arrows correspond
to dressed anomalous propagators $\widetilde{F}$ and
$\widetilde{F}^{+}$ (\ref{2.4}).} \label{Fig3}
\end{figure}

Let us embed impurities into a above described metallic matrix.
The impurities interact with conduction electrons of the metal in
a retarding manner. In the simplest case the impurity is modelled
with the oscillator (\ref{1.9}) with eigenfrequency $\omega_{0}$.
As it is well known Gor'kov equation for a dirty superconductor
has a form shown in Fig.\ref{Fig3}. The diagrams are analogous to
diagrams of a dirty superconductor with elastic impurities
(Apendix \ref{A}). Their sense is that electrons pair in the
metallic matrix at first, then normal and anomalous propagators
are dressed by interaction with impurities \citep{abrik}. However
now lines of the interaction (dotted lines) transfer energy. We
neglect cross diagrams (the diagrams like in Fig.\ref{Fig2}b). For
elastic scattering a small parameter of contribution of the cross
diagrams is $\frac{1}{k_{F}l}\ll 1$ \citep{sad}. For a retarded
interaction with impurities the small parameter is a ratio
(\ref{1.19}) $\frac{\omega_{0}}{\varepsilon_{F}}\ll 1$. As in
\citep{sad} we suppose the order parameter is self—averaging:
$\langle\Delta^{2}(\textbf{r})\rangle-\langle\Delta(\textbf{r})\rangle^{2}=0$.
This means to neglect scattering of Cooper pairs by fluctuations
of the gap, it is valid at $\frac{1}{k_{F}l}\ll 1$. In an analytic
form the equations in Fig.\ref{Fig3} are:
\begin{equation}\label{2.3}
    \left\{\begin{array}{c}
      \widetilde{G}(i\widetilde{\varepsilon}_{n}-\xi)+\widetilde{F}^{+}\widetilde{\Delta}=i \\
      \\
      \widetilde{F}^{+}(i\widetilde{\varepsilon}_{n}+\xi)+\widetilde{G}\widetilde{\Delta}^{+}=0 \\
    \end{array}\right\}
\end{equation}
Solutions of the set of equations are normal and anomalous
propagators:
\begin{equation}\label{2.4}
    \widetilde{G}(\varepsilon_{n},\xi)=i\frac{i\widetilde{\varepsilon}_{n}+\xi}
    {(i\widetilde{\varepsilon}_{n})^{2}-\xi^{2}-|\widetilde{\Delta}(\varepsilon_{n})|^{2}},\quad
    \widetilde{F}^{+}(\widetilde{\varepsilon}_{n},\xi)=\frac{-i\widetilde{\Delta}^{+}(\varepsilon_{n})}
    {(i\widetilde{\varepsilon}_{n})^{2}-\xi^{2}-|\widetilde{\Delta}(\varepsilon_{n})|^{2}},
\end{equation}
where a renormalized gap $\widetilde{\Delta}$ and a renormalized
energy parameter $\widetilde{\varepsilon}_{n}$ are determined with
equations (here $\varepsilon_{n}=(2n+1)\pi T$ and
$\varepsilon_{n'}=(2n'+1)\pi T$):
\begin{eqnarray}
\widetilde{\Delta}^{+}(\varepsilon_{n})&=&\Delta^{+}(\varepsilon_{n})+\rho\int\frac{d^{3}p}{(2\pi)^{3}}\left|\upsilon(\textbf{k}-\textbf{p})\right|^{2}
T\sum_{n'=-\infty}^{+\infty}\left(-i\widetilde{F}^{+}(\textbf{p},\varepsilon_{n'})\right)iD(\varepsilon_{n}-\varepsilon_{n'})\label{2.5}\\
&=&\Delta^{+}(\varepsilon_{n})+\rho\left|\upsilon\right|^{2}\nu_{F}\frac{2}{\omega_{0}}
\sum_{n'=-\infty}^{+\infty} \frac{\pi
T\widetilde{\Delta}^{+}(\varepsilon_{n'})}{\sqrt{\widetilde{\varepsilon}_{n'}^{2}(\varepsilon_{n'})+|\widetilde{\Delta}(\varepsilon_{n'})|^{2}}}
\frac{\omega_{0}^{2}}{(\varepsilon_{n}-\varepsilon_{n'})^{2}+\omega_{0}^{2}}
\nonumber\\
i\widetilde{\varepsilon}_{n}(\varepsilon_{n})&=&i\varepsilon_{n}+\rho\int\frac{d^{3}p}{(2\pi)^{3}}\left|\upsilon(\textbf{k}-\textbf{p})\right|^{2}
T\sum_{n'=-\infty}^{+\infty}i\widetilde{G}(\textbf{p},\varepsilon_{n'})iD(\varepsilon_{n}-\varepsilon_{n'})\label{2.6}\\
&=&i\varepsilon_{n}+\rho\left|\upsilon\right|^{2}\nu_{F}\frac{2}{\omega_{0}}
\sum_{n'=-\infty}^{+\infty} \frac{\pi
Ti\widetilde{\varepsilon}_{n'}(\varepsilon_{n})}{\sqrt{\widetilde{\varepsilon}_{n'}^{2}(\varepsilon_{n'})+|\widetilde{\Delta}(\varepsilon_{n'})|^{2}}}
\frac{\omega_{0}^{2}}{(\varepsilon_{n}-\varepsilon_{n'})^{2}+\omega_{0}^{2}}.\nonumber
\end{eqnarray}
Since the interaction with an impurity is short-range then we can
suppose
$\upsilon(\textbf{k}-\textbf{p})\approx\upsilon=\texttt{const}$.
Since spectrum of quasiparticles near Fermi surface is linear then
the integration over momentums can be simplified:
$\frac{d^{3}p}{(2\pi)^{3}}\approx\nu_{F}d\xi$. The gap $\Delta$ is
determined by the same equation (\ref{2.1}) however with a dressed
anomalous propagator $\widetilde{F}$ from Eq.(\ref{2.4}):
\begin{eqnarray}\label{2.7}
\Delta^{+}(\varepsilon_{n})=gT\sum_{n'=-\infty}^{+\infty}\int_{-\infty}^{+\infty}
d\xi(-i)\widetilde{F}^{+}(\varepsilon_{n'},\xi)w_{\omega_{\texttt{D}}}(\varepsilon_{n},\varepsilon_{n'})
\end{eqnarray}
Our problem is to calculate the parameters $\widetilde{\Delta}$
and $\widetilde{\varepsilon}_{n}$, and with them to calculate
critical temperature $\Delta(T_{\texttt{С}})=0$ in a system metal
matrix+impurities. It should be noted that these functions are
determined self-consistently, that is any corrections to their
values in a pure metal $\Delta,\varepsilon_{n}$ are determined by
the sought quantities
$\widetilde{\Delta},\widetilde{\varepsilon}_{n}$. In a general
case to solve the set of equations (\ref{2.5},\ref{2.6},\ref{2.7})
is very problematically. We will consider some limit cases and
approximations.

\subsection{A limit case of high temperature $T\gg\omega_{0}$.}\label{Anderson2}

In a case $T\gg\omega_{0}$ we can neglect an energy transfer along
the lines of interaction with an impurity because
$(\varepsilon_{n}-\varepsilon_{n'})^{2}=4\pi^{2}T^{2}(n-n')^{2}\gg\omega_{0}^{2}$,
if $n'\neq n$. That is we have
\begin{equation}\label{2.8}
    \frac{\omega_{0}^{2}}{(\varepsilon_{n}-\varepsilon_{n'})^{2}+\omega_{0}^{2}}\rightarrow
    \begin{array}{cc}
      \frac{\omega_{0}^{2}}{4\pi^{2}T^{2}(n-n')^{2}}\rightarrow0 & \texttt{for}\quad n'\neq n \\
      \\
      1 & \texttt{for}\quad n'=n \\
    \end{array}
\end{equation}
Thus in the sum $\sum_{n'=-\infty}^{+\infty}$ terms with $n'=n$
survive only. Then Eqs.(\ref{2.5},\ref{2.6}) are reduced to
\begin{eqnarray}\label{2.9}
&&\widetilde{\Delta}^{+}=\Delta^{+}+\frac{2\pi\rho\left|\upsilon\right|^{2}\nu_{F}T}{\omega_{0}}
\frac{\widetilde{\Delta}^{+}}{\sqrt{\widetilde{\varepsilon}_{n'}^{2}+|\widetilde{\Delta}|^{2}}}
\nonumber\\
\\
&&i\widetilde{\varepsilon}_{n}=i\varepsilon_{n}+\frac{2\pi\rho\left|\upsilon\right|^{2}\nu_{F}T}{\omega_{0}}
\frac{i\widetilde{\varepsilon}_{n}}{\sqrt{\widetilde{\varepsilon}_{n'}^{2}+|\widetilde{\Delta}|^{2}}}.\nonumber
\end{eqnarray}
We can see that the limit $T\gg\omega_{0}$ corresponds to elastic
scattering by impurities with a scattering rate
$\frac{1}{2\pi\tau}=\rho\upsilon^{2}\nu_{F}\frac{2T}{\omega_{0}}$.
Solving Eq.(\ref{2.9}) we find that the gap and the energetic
parameter are renormalized similarly:
\begin{equation}\label{2.10}
\frac{\widetilde{\Delta}}{\Delta}=\frac{\widetilde{\varepsilon_{n}}}{\varepsilon_{n}}=1+\frac{1}{2\tau}\frac{1}{\sqrt{\varepsilon_{n}^{2}+\Delta^{2}}}.
\end{equation}
The relation (\ref{2.10}) means realization of Anderson's theorem
(the gap and, accordingly, critical temperature do not change):
\begin{equation}\label{2.11}
\Delta=g\pi
T\sum_{n=-\infty}^{+\infty}\frac{\widetilde{\Delta}}{\sqrt{\widetilde{\varepsilon}_{n'}^{2}+\widetilde{\Delta}^{2}}}=
g\pi
T\sum_{n=-\infty}^{+\infty}\frac{\Delta}{\sqrt{\varepsilon_{n'}^{2}+\Delta^{2}}}.
\end{equation}
Thus at temperatures being much more than oscillation frequency of
impurities $\omega_{0}$ the scattering by the impurities
influences trivially upon supervonductive properties of a metal:
it does not change a critical temperature and it reduces a
coherence length $\frac{1}{\xi}=\frac{1}{\xi_{0}}+\frac{1}{l}$.

Let us consider a special case $\omega_{0}=0$. Then in the sums
$\sum_{n'=-\infty}^{+\infty}$ in the Eqs.(\ref{2.5},\ref{2.6}) the
diagonal term $n=n'$ survives only, where we have an uncertainty
$\frac{0}{0}$. Therefore in this case an elastic scattering takes
place because an energy parameter is not transferred. Let us
reveal the uncertainty. In a formula (\ref{1.9}) if
$\omega_{0}\rightarrow 0$ we have
$\upsilon(\textbf{q},t')\upsilon(-\textbf{q},t'')=|\upsilon(\textbf{q})|^{2}$.
The term
$\left|\upsilon\right|^{2}\frac{2\omega_{0}}{(\omega_{m})^{2}+\omega_{0}^{2}}\equiv
U(\omega_{m})$ appears due Fourier transition \cite{matt}:
\begin{eqnarray}
U(\omega_{m})=\frac{1}{2}\int_{-1/T}^{1/T}\exp\left(i\omega_{m}\tau\right)\upsilon(\tau')\upsilon(\tau'')d\tau,\nonumber
\end{eqnarray}
where $\tau=\tau''-\tau'\in(-1/T,+1/T)$ (thermodynamic time),
$\omega_{m}=\pi mT$, $m$ are integers (excitations of an impurity
are bosons). If
$\upsilon(\tau')\upsilon(\tau'')=\upsilon^{2}=\texttt{const}$,
then $\lim_{m\rightarrow 0}U(\omega_{m})=1/T$. The limit $m=0$
corresponds to above-mentioned case $n=n'$. This means that
\begin{eqnarray}
\lim_{\omega_{0}\rightarrow 0}\rho\left|\upsilon\right|^{2}\nu_{F}
\sum_{n'=-\infty}^{+\infty} \frac{\pi
T\widetilde{\Delta}^{+}(\varepsilon_{n'})}{\sqrt{\widetilde{\varepsilon}_{n'}^{2}(\varepsilon_{n'})+|\widetilde{\Delta}(\varepsilon_{n'})|^{2}}}
\frac{2\omega_{0}}{(\varepsilon_{n}-\varepsilon_{n'})^{2}+\omega_{0}^{2}}=
\rho\left|\upsilon\right|^{2}\nu_{F} \frac{\pi
\widetilde{\Delta}^{+}(\varepsilon_{n})}{\sqrt{\widetilde{\varepsilon}_{n}^{2}(\varepsilon_{n})+|\widetilde{\Delta}(\varepsilon_{n})|^{2}}}
\nonumber
\end{eqnarray}
That is we have an elastic scattering in the limit of impurities
without internal structure.

\subsection{A self-consistent approximate solution for a general case.}

From Eqs.(\ref{2.5},\ref{2.6}) we can see that \textit{in a
general case} \emph{the gap and the energy parameter are
renormalized differently due a retarded interaction with
impurities. Hence Anderson's theorem is violated.} This is a
result of the fact that the equations  include a summation over
the energy parameter (an index $n'$) unlike an elastic case, and
under the summation sign different functions are -
$\widetilde{\Delta}$ in the equation (\ref{2.5}) (even) and
$\widetilde{\varepsilon}_{n}$ in the equation (\ref{2.6}) (odd).
The set of equations (\ref{2.5},\ref{2.6}) can be simplified using
an approximation of an electron-electron interaction amplitude
$gw(\varepsilon_{n},\varepsilon_{n'})$ with a method stated in
\citep{levit}:
\begin{equation}\label{2.12}
gw(\varepsilon_{n},\varepsilon_{n'})=
g\frac{\omega^{2}}{(\varepsilon_{n}-\varepsilon_{n'})^{2}+\omega^{2}}
\longrightarrow
gw(\varepsilon_{n})w(\varepsilon_{n'})=g\frac{\omega}{\sqrt{\varepsilon_{n}^{2}+\omega^{2}}}
\frac{\omega}{\sqrt{\varepsilon_{n'}^{2}+\omega}}.
\end{equation}
Here $\omega=\omega_{\texttt{D}}, \omega_{0},\ldots$ is
characteristic frequency of the interaction. The approximation
(\ref{2.12}) corresponds to separation of a contribution of terms
with $n'\neq n$ because at $T\gg\omega_{0}$ we have a limit:
\begin{eqnarray}
\frac{\pi
T}{\omega_{0}}\frac{\omega_{0}^{2}}{\varepsilon_{n}^{2}+\omega_{0}^{2}}=
\frac{\pi T}{\omega_{0}}\frac{(\omega_{0}/\pi
T)^{2}}{(2n+1)^{2}+(\omega_{0}/\pi T)^{2}}\longrightarrow
\frac{\omega_{0}}{\pi T}\frac{1}{(2n+1)^{2}}\rightarrow 0\nonumber
\end{eqnarray}
The terms with $n'=n$ in sums (\ref{2.5},\ref{2.6}) neglected in
the the approximation (\ref{2.12}) do not influence upon a gap and
a critical temperature. The terms describe a scattering of
electrons by thermal oscillations of an impurity. The thermal
oscillations behave like static impurities with effective
concentration $\rho\frac{2T}{\omega_{0}}$. The scattering gives an
additional contribution in resistance of the metal analogously to
a contribution of thermal phonons. Thus the terms with $n'\neq n$
can violate Anderson's theorem only.

Due the approximation (\ref{2.12}) Eqs.(\ref{2.5},\ref{2.6}) take
forms:
\begin{eqnarray}
&&\widetilde{\Delta}^{+}(\varepsilon_{n})=\Delta^{+}(\varepsilon_{n})+\rho\left|\upsilon\right|^{2}\nu_{F}\frac{2}{\omega_{0}}
\sum_{n'=-\infty}^{+\infty} \frac{\pi
T\widetilde{\Delta}^{+}(\varepsilon_{n'})}{\sqrt{\widetilde{\varepsilon}_{n'}^{2}+|\widetilde{\Delta}(\varepsilon_{n'})|^{2}}}
w(\varepsilon_{n})w(\varepsilon_{n'})
\label{2.13}\\
&&\widetilde{\varepsilon}_{n}=\varepsilon_{n}+\rho\left|\upsilon\right|^{2}\nu_{F}\frac{2}{\omega_{0}}
\sum_{n'=-\infty}^{+\infty} \frac{\pi
T\widetilde{\varepsilon}_{n'}}{\sqrt{\widetilde{\varepsilon}_{n'}^{2}+|\widetilde{\Delta}(\varepsilon_{n'})|^{2}}}
\frac{\omega_{0}}{\sqrt{\varepsilon_{n}^{2}+\omega_{0}^{2}}}
\frac{\omega_{0}}{\sqrt{\varepsilon_{n'}^{2}+\omega_{0}}}=\varepsilon_{n}+0.\label{2.14}
\end{eqnarray}
In Eq.(\ref{2.14}) in the second term under the summation sign a
odd function of $n'$ is, hence the energy parameter is not
renormalized $\widetilde{\varepsilon}_{n}=\varepsilon_{n}$. Let us
consider a case when Debye frequency of the matrix and a frequency
of impurities coincide: $\omega_{\texttt{D}}=\omega_{0}$. A
dependence of the gap on energy we can write in a form
$\Delta(\varepsilon_{n})=\Delta
w_{\omega_{\texttt{D}}}(\varepsilon_{n})=\Delta
w_{\omega_{0}}(\varepsilon_{n})$ and
$\widetilde{\Delta}(\varepsilon_{n})=\Delta
w_{\omega_{0}}(\varepsilon_{n})$ (as in Eq.(\ref{2.2a})). Then
Eq.(\ref{2.13}) is simplified:
\begin{eqnarray}\label{2.13a}
\widetilde{\Delta}^{+}=\Delta^{+}+\rho\left|\upsilon\right|^{2}\nu_{F}\frac{2}{\omega_{0}}
\sum_{n'=-\infty}^{+\infty} \frac{\pi
T\widetilde{\Delta}^{+}}{\sqrt{\varepsilon_{n'}^{2}+|\widetilde{\Delta}|^{2}w^{2}(\varepsilon_{n'})}}
w^{2}(\varepsilon_{n'})
\end{eqnarray}
The equation can be rewritten in a form:
\begin{eqnarray}\label{2.15}
&&\widetilde{\Delta}^{+}=\frac{\Delta^{+}}{1-\rho\left|\upsilon\right|^{2}\nu_{F}\frac{2}{\omega_{0}}
\sum_{n=-\infty}^{+\infty}
\frac{1}{\sqrt{(2n+1)^{2}+\left(\widetilde{\Delta}/\pi
T\right)^{2}w^{2}(n)}}\frac{\left(\omega_{0}/\pi
T\right)^{2}}{(2n+1)^{2}+\left(\omega_{0}/\pi T\right)^{2}}}
\end{eqnarray}
and it can be solved for $\widetilde{\Delta}$. Obtained solution
has to be substituted into Eq.(\ref{2.7}) determining the gap
$\Delta(T)$ in a system matrix+impurities. In a case
$\omega_{\texttt{D}}\neq\omega_{0}$ Eq.(\ref{2.13}) can be reduced
to a form:
\begin{eqnarray}\label{2.15a}
&&\widetilde{\Delta}^{+}(\varepsilon_{n})=\Delta^{+}
w_{\omega_{\texttt{D}}}(\varepsilon_{n})+\Delta^{+}
w_{\omega_{0}}(\varepsilon_{n})\frac{f}{1-h},
\end{eqnarray}
where functions $f$ and $h$ are
\begin{eqnarray}\label{2.15b}
&&f=\rho\left|\upsilon\right|^{2}\nu_{F}\frac{2}{\omega_{0}}
\sum_{n=-\infty}^{+\infty}
\frac{1}{\sqrt{(2n+1)^{2}+\left(\widetilde{\Delta}(\varepsilon_{n})/\pi
T\right)^{2}}}\frac{\left(\omega_{\texttt{D}}/\pi
T\right)}{\sqrt{(2n+1)^{2}+\left(\omega_{\texttt{D}}/\pi
T\right)^{2}}}\frac{\left(\omega_{0}/\pi
T\right)}{\sqrt{(2n+1)^{2}+\left(\omega_{0}/\pi T\right)^{2}}}\nonumber\\
&&h=\rho\left|\upsilon\right|^{2}\nu_{F}\frac{2}{\omega_{0}}
\sum_{n=-\infty}^{+\infty}
\frac{1}{\sqrt{(2n+1)^{2}+\left(\widetilde{\Delta}(\varepsilon_{n})/\pi
T\right)^{2}}}\frac{\left(\omega_{0}/\pi
T\right)^{2}}{(2n+1)^{2}+\left(\omega_{0}/\pi T\right)^{2}}
\end{eqnarray}
If $\omega_{\texttt{D}}=\omega_{0}$ we have $h=f$, hence
$\widetilde{\Delta}^{+}(\varepsilon_{n})=\widetilde{\Delta}^{+}
w_{\omega_{0}}=\Delta^{+} w_{\omega_{0}}\frac{1}{1-h}$ that
coincides Eq.(\ref{2.15}).

\subsubsection{Critical temperature.}

\begin{figure}[ht]
\includegraphics[width=8.5cm]{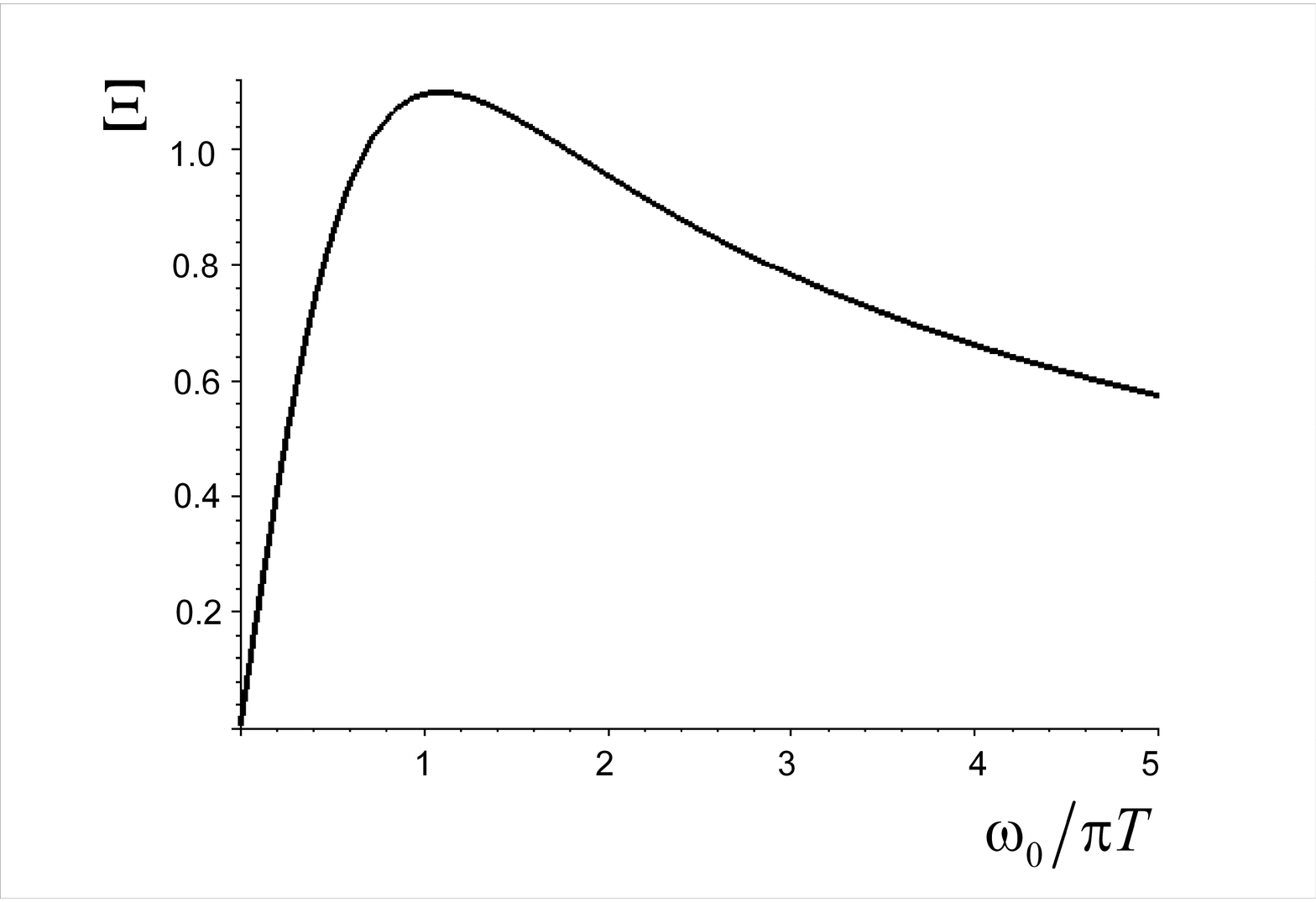}
\caption{The effectiveness function $\Xi\left(\omega_{0}/T\right)$
as a function of the ratio of an impurity's oscillation frequency
to temperature.} \label{Fig4}
\end{figure}

The problem is essentially simplified if we find critical
temperature only. Then
$\widetilde{\Delta}(T_{\texttt{C}})=\Delta(T_{\texttt{C}})=0$ and
Eq.(\ref{2.15a}) is reduced to a form:
\begin{equation}\label{2.16a}
 \widetilde{\Delta}^{+}(\varepsilon_{n})=\Delta^{+}
w_{\omega_{\texttt{D}}}(\varepsilon_{n})+\Delta^{+}
w_{\omega_{0}}(\varepsilon_{n})\frac{\frac{2\rho\left|\upsilon\right|^{2}\nu_{F}}{\pi
T}\Upsilon\left(\frac{\omega_{\texttt{D}}}{\pi
T},\frac{\omega_{0}}{\pi T}\right)}
{1-\frac{2\rho\left|\upsilon\right|^{2}\nu_{F}}{\pi
T}\Xi\left(\frac{\omega_{0}}{\pi T}\right)},
\end{equation}
where $\Xi$ is a function which we name an \emph{effectiveness
function} (Fig.\ref{Fig4}):
\begin{equation}\label{2.17}
\Xi\left(\frac{\omega_{0}}{\pi T}\right)=\frac{\pi
T}{\omega_{0}}\left[\gamma+2\texttt{ln}2+
\frac{1}{2}\Psi\left(\frac{1}{2}-\frac{i}{2}\frac{\omega_{0}}{\pi
T}\right)+\frac{1}{2}\Psi\left(\frac{1}{2}+\frac{i}{2}\frac{\omega_{0}}{\pi
T}\right)\right].
\end{equation}
Here $\Psi$ is a digamma function , $\gamma\approx 0.577$ is Euler
constant. \emph{The effectiveness function describes an influence
of the impurities upon a superconductor depending on their
oscillation frequency} $\omega_{0}$. \emph{The effectiveness
function} $\Xi\left(\omega_{0}, T_{\texttt{C}}^{\ast}\right)$
\emph{determines such an oscillation frequency of an impurity}
\emph{to get the critical temperature} $T_{\texttt{C}}^{\ast}$
\emph{with the least concentration of the impurities}. On the one
hand the lower frequency the stronger scattering of electrons by
the impurities
$\rho\left|\upsilon\right|^{2}\nu_{F}\frac{2}{\omega_{0}}$,
because it is necessary less expenditure of electron's energy to
"swing" an oscillator. On the other hand at temperatures
$T\gtrsim\omega_{0}$ a thermal noise destroys changes of
oscillators' states by electrons. Thus in a region of frequencies
and temperatures $T\gg\omega_{0}$ the impurities' effectiveness
falls $\Xi\left(\frac{\omega_{0}}{T}\rightarrow
0\right)\rightarrow\frac{7}{4}\zeta(3)\frac{\omega_{0}}{\pi
T}\rightarrow 0$. This result is in agreement with a result in
Subsection \ref{Anderson2}. At $\omega_{0}\gg T$ the oscillations
are "freezed" and energy level transitions are determined by an
interaction with metal's electrons only. However for too large
frequencies $\omega\sim\varepsilon_{F}$ the interaction is weak.
Hence in a region of large frequencies $\omega_{0}$ the
effectiveness is slowly decreasing:
$\Xi\left(\frac{\omega_{0}}{T}\rightarrow
\infty\right)\rightarrow\frac{\pi
T}{\omega_{0}}\texttt{ln}\left(\frac{2}{\gamma}\frac{\omega_{0}}{\pi
T}\right)\rightarrow 0$. An optimal value of the oscillation
frequency is $\frac{\omega_{0}}{\pi T}=1.09$ at given temperature
when the effectiveness function reaches its maximum value
$\Xi_{\texttt{max}}=1.10$ (Fig.\ref{Fig4}). A function $\Upsilon$
is analogous to the effectiveness function $\Xi$ with the
difference that it depends on both a matrix's frequency
$\omega_{\texttt{D}}$ and an impurity's frequency $\omega_{0}$.
However it does not play a principal role because it is in the
numerator. If the frequencies are equal
$\omega_{\texttt{D}}=\omega_{0}$ then Eq.(\ref{2.16a}) passes into
Eq.(\ref{2.15}).

Now let us consider a multiplier
$\frac{2\rho\left|\upsilon\right|^{2}\nu_{F}}{\pi T}$ in the
formula (\ref{2.16a}). By analogy with the elastic scattering
(Appendix \ref{A}) the value
$\frac{1}{\tau}=2\pi\rho\upsilon^{2}\nu_{F}$ can be considered as
a scattering rate. Then a free length is
\begin{equation}\label{2.18}
    l=v_{F}\tau=\frac{v_{F}}{2\pi\rho\upsilon^{2}\nu_{F}}=\frac{k_{F}}{2\pi\rho\upsilon^{2}\nu_{F}m}
\end{equation}
then
\begin{equation}\label{2.19}
\frac{2\rho\left|\upsilon\right|^{2}\nu_{F}}{\pi
T}\equiv\frac{1/\tau}{\pi^{2}T}=\frac{2\varepsilon_{F}}{\pi^{2}T}\frac{1}{k_{F}l}.
\end{equation}
It should be noted that real values of a ratio of a reverse free
length $1/l$ to Fermi momentum of a matrix $k_{F}$ (which is equal
to reverse interatomic distance $k_{F}\sim 1/a$) is
$\frac{1}{k_{F}l}\sim \frac{a}{l}\ll 1$. When
$\frac{1}{k_{F}l}\gtrsim 1$ a transition in a state of Anderson
insulator can take place \citep{sad,sad1} (a localization with
impurities). However if the scattering is essentially inelastic
then the transition can be suppressed.

\begin{figure}[ht]
\includegraphics[width=8.5cm]{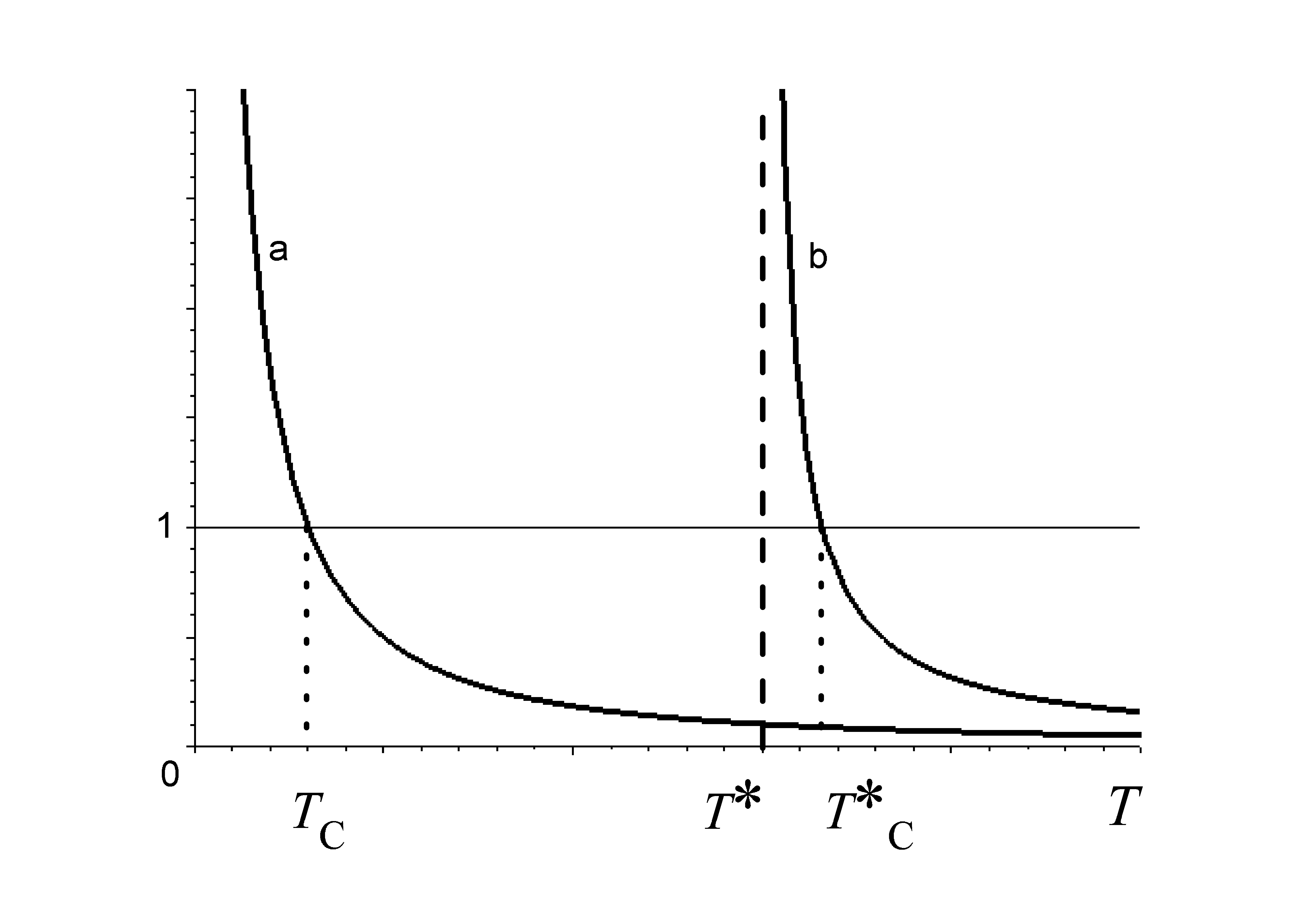}
\caption{A graphic representation of the equations for the
critical temperature. The curve (a) corresponds to Eq.(\ref{2.1})
and determines the critical temperature of a pure superconductor
$T_{\texttt{C}}$. The curve (b) corresponds to Eq.(\ref{2.20}) and
determines critical temperature $T_{\texttt{C}}^{\ast}$ of a
system metal+impurities. In a point $T=0$ the equation for a pure
metal has a singularity. For the system metal+impureties the
singularity exists at nonzero temperature $T^{\ast}$.}
\label{Fig5}
\end{figure}

Substituting the renormalized gap (\ref{2.16a})
$\widetilde{\Delta}$ in Eq.(\ref{2.7}) and using the
approximations (\ref{2.2a}) and (\ref{2.12}) we obtain an equation
to find critical temperature:
\begin{eqnarray}\label{2.20}
&&\Delta^{+}(\varepsilon_{n})=gT\sum_{n'=-\infty}^{+\infty}
\frac{\pi\widetilde{\Delta}^{+}(\varepsilon_{n})}{|\varepsilon_{n'}|}
w_{\omega_{\texttt{D}}}(\varepsilon_{n},\varepsilon_{n'})\nonumber\\
&&\Rightarrow 1=g\sum_{n'=-\infty}^{+\infty} \frac{\pi
T}{|\varepsilon_{n'}|}\left[w_{\omega_{\texttt{D}}}^{2}(\varepsilon_{n'})+
w_{\omega_{\texttt{D}}}(\varepsilon_{n'})w_{\omega_{0}}(\varepsilon_{n'})
\frac{\frac{2\varepsilon_{F}}
{\pi^{2}T}\frac{1}{k_{F}l}\Upsilon\left(\frac{\omega_{\texttt{D}}}{\pi
T},\frac{\omega_{0}}{\pi T}\right)} {1-\frac{2\varepsilon_{F}}
{\pi^{2}T}\frac{1}{k_{F}l}\Xi\left(\frac{\omega_{0}}{\pi
T}\right)} \right]
\end{eqnarray}
In a limit $l\rightarrow\infty$ the equation (\ref{2.20}) pass
into equation (\ref{2.1}) for a pure superconductor. Graphicly
Eq.(\ref{2.20}) is shown in Fig.\ref{Fig5}. The critical
temperature is determined by an intersection  with the line 1 and
the second term of Eq.(\ref{2.20}) as a function of temperature.
The curve (a) determines critical temperature of a pure
superconductor $T_{\texttt{C}}$. The more a coupling constant $g$
the more $T_{\texttt{C}}$. The curve (b) determines critical
temperature $T_{\texttt{C}}^{\ast}$ of a system metal+impurities.
The temperature $T^{\ast}$ is determined by a zero in the
denominator in the formula (\ref{2.20}):
\begin{equation}\label{2.21}
\frac{2\varepsilon_{F}}{\pi^{2}T^{\ast}}\frac{1}{k_{F}l}\Xi\left(\frac{\omega_{0}}{\pi
T^{\ast}}\right)=1\Longleftrightarrow
\frac{1/\tau}{\pi^{2}T^{\ast}}\Xi\left(\frac{\omega_{0}}{\pi
T^{\ast}}\right)=1.
\end{equation}
Moreover we can see that an inequality
$T_{\texttt{C}}^{\ast}\gtrsim T^{\ast}>T_{\texttt{C}}$ takes
place. In a point $T=T^{\ast}$ the second member in
Eq.(\ref{2.20}) is singular $g\cdot\infty$. \emph{The singularity
appears due a combined consistent pairing action of matrix's
phonons and impurities' oscillations on electrons on the
assumption of the averaging over a disorder} (\ref{1.6}),
\emph{with a correlator "white noise"} (\ref{1.7}). \emph{The
intensification of the pairing has sense in the presence of
electron-electron attraction in a matrix only}
$g=\lambda-\mu^{\ast}>0$. \emph{Thus the impurities play a role of
a catalyst of superconductivity.} The singularity temperature
$T^{\ast}$ is determined by electronic parameters of a matrix and
a coupling constant with impurities $\rho\upsilon^{2}\nu_{F}$.
However $T^{\ast}$ does not depend on a frequency of a pairing
interaction in a matrix $\omega_{D}$ if only it is nonzero and the
coupling constant $g$ if only it corresponds to attraction
$g=\lambda-\mu^{\ast}>0$. Moreover it is necessary to notice that
$T_{\texttt{C}}^{\ast}$ differs from $T^{\ast}$ little in
consequence of a dependence of interaction with impurities on
temperature $\frac{1}{T^{\ast}}\Xi\left(\frac{\omega_{0}}{\pi
T^{\ast}}\right)$.  Therefore the temperature $T^{\ast}$ can be
used as a lower estimate of the critical temperature. Pure
superconductors have the singularity temperature too, however it
equals to zero always $T^{\ast}_{l=\infty}=0$ (Fig.\ref{Fig5}).

\begin{figure}[ht]
\includegraphics[width=8.0cm]{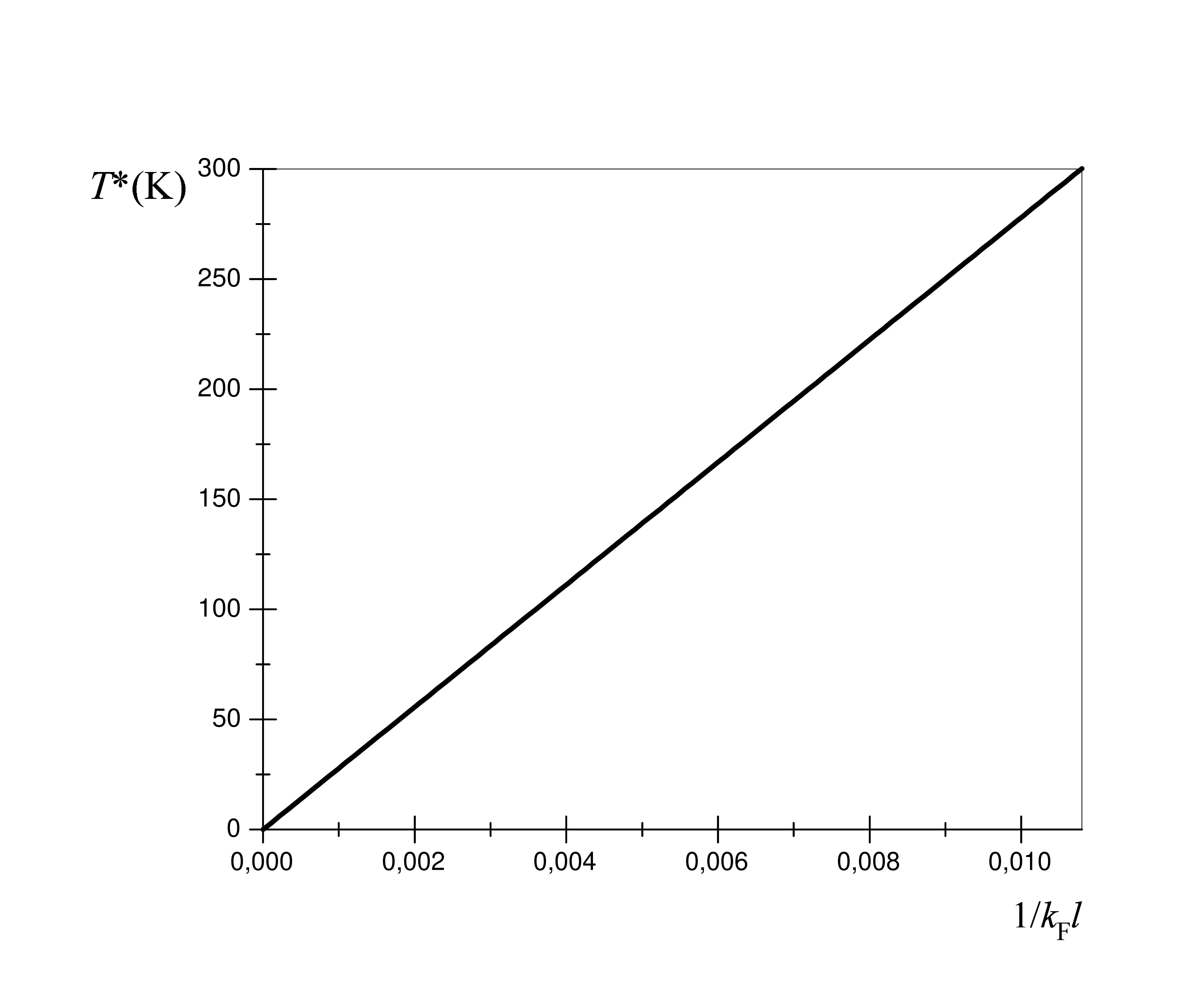}
\caption{The singularity temperature $T^{\ast}$ as a function of a
parameter $\frac{1}{k_{F}l}$ for a matrix $\texttt{Al}$ providing
that the oscillation frequency of impurities is in optimal ratio
to the temperature $\frac{\omega_{0}}{\pi T^{\ast}}\approx 1$.}
\label{Fig6}
\end{figure}

\emph{Estimation of the critical temperature}
$T_{\texttt{C}}^{\ast}$ \emph{states that it can essentially
exceed critical temperature of a corresponding pure metal on the
assumption of optimal choice of parameters of the matrix and the
impurities.} For example, a pure crystal of $\texttt{Al}$ has
critical temperature $T_{\texttt{C}}=1.2\texttt{K}$ corresponding
to parameters $g=0.17$, $\omega_{\texttt{D}}=375\texttt{K}$. Fermi
energy and velocity are equal to $\varepsilon_{F}=13.6\cdot
10^{4}\texttt{K}$, $v_{F}=2.03\cdot 10^{6}\texttt{m/s}$
accordingly. Let the oscillation frequency is chosen in the
optimal ratio to a desired temperature $\frac{\omega_{0}}{\pi
T^{\ast}}\approx 1$, that is a value of the effectiveness function
is $\Xi=1\approx\Xi_{\texttt{max}}$. Then we can plot the
singularity temperature $T^{\ast}$ as a function of a parameter
$\frac{1}{k_{F}l}$ - Fig.\ref{Fig6}. In the figure we can see that
if the parameter is $\frac{1}{k_{F}l}\approx 0.01\ll 1$ the
singularity temperature reaches giant (room) values $T^{\ast}\sim
300\texttt{K}$ in comparison with critical temperature of the pure
metal. It should be noted that in Fig.(\ref{Fig6}) \textit{various
values of $l$ and, accordingly, various values of $T^{\ast}$
correspond to various impurities chosen so that the frequency is
in optimal ratio to the temperature} $\frac{\omega_{0}}{\pi
T^{\ast}}\approx 1$. With help an expression (\ref{2.19}) we can
calculate that to reach the critical temperature $\sim
300\texttt{K}$ the free length must be $l\approx 12a$ where
lattice constant of $\texttt{Al}$ is $a=4.08\texttt{A}$.

It should be noted we cannot infinitely increase
$T_{\texttt{C}}^{\ast}$ decreasing $l$. To decrease $l$ we must to
increase impurity's concentration $\rho$. However this means the
impurities replace atoms of metals. This results in decreasing of
concentration of conduction electrons $n$ and, in turn, in
decreasing of density of states on Fermi surface
$\nu_{F}=\frac{mk_{F}}{2\pi^{2}}=\frac{m(3\pi^{2}n)^{1/3}}{2\pi^{2}}$
figuring in Eqs.(\ref{2.5},\ref{2.6}) and in
Eqs.(\ref{2.1},\ref{2.7}) ($\nu_{F}$ is included into a coupling
constant $g$). If the impurity's concentration reaches atoms'
concentration in a pure metal $N_{0}/V$ then
$T_{\texttt{C}}^{\ast}=0$ because electrons' concentration becomes
zero. Moreover, for large concentrations of impurities the concept
of impurities and metal is senseless, and much earlier than
$\rho\sim N_{0}/V$ the metal can be destroyed. We can also
increase electron-impurity coupling constant $\upsilon$. However
large value of $\upsilon$ is unphysical (as the electron-phonon
coupling constant) and its calculation requires special
consideration.

\paragraph{The gap.}

To solve the self-consistent equations (\ref{2.15}) or
(\ref{2.15a}) for the gap $\widetilde{\Delta}$ at temperature
$T<T^{\ast}_{\texttt{C}}$ is a more difficult problem than the
previous one for the critical temperature. In Eq.(\ref{2.20}) for
the critical temperature we have the singularity at temperature
$T^{\ast}< T^{\ast}_{\texttt{C}}$ determined by Eq.(\ref{2.21})
(zero in the denominator). However at presence of a nonzero gap
$\widetilde{\Delta}$ the singularity is absent in consequence of
self-consistency of the equations. Hence at any temperature
$T<T^{\ast}_{\texttt{C}}$ the order parameter $\Delta(T)$
determined by Eq. (\ref{2.7}) for a system matrix+impurities is a
finite value.

\section{Discussion.}\label{concl}

In this article a theory of disordered metals is generalized if a
retarded interaction of conduct electrons with impurities takes
place. In consequence of averaging over the disorder we have
diagram rules to be analogous to diagram rules for scattering by
elastic impurities, however the lines of interaction carry both
momentum and energy. In a basic approximation an impurity is a
harmonic oscillator with some frequency $\omega_{0}$. As a result
of correlations between successive scatterings we have a picture
that as though ''collective excitations'' propagates through the
system, and scattering of metal's quasi-particles by impurities is
determined by some coupling constant which depends on
concentration of the impurities. As a result of inelasticity of
the scattering by impurities the quantum contribution to
conductivity (localization) can be suppressed. If temperature is
much greater than the oscillation frequency of impurities
$T/\omega_{0}\gg 1$ then the scattering by the impurities can be
considered as elastic scattering by impurities with effective
concentration $\rho\frac{2T}{\omega_{0}}$ (if only $\omega_{0}\neq
0$).

Injection of the impurities into three-dimension $s$-wave
superconductor essentially influences on its superconductive
properties. A gap and an energy parameter are renormalizated
differently due a retarded interaction of metal's quasi-particles
with the impurities. Mechanism of influence of an impurity on
Cooper pair consists of the following: the first electron changes
impurity's state, then the second one interacts with the impurity
changed by the first electron, thus a correlation between the
electrons appears that increases their binding energy. This means
that of amplification of superconductive properties is a result of
the effects of memory in the scattering by an impurity. This
mechanism causes violation of Anderson's theorem in the direction
of increasing of critical temperature. Influence of impurities
upon the critical temperature determined by a reverse free length
$1/l$ or a scattering rate $1/\tau=v_{F}/l$. The critical
temperature essentially depends on the oscillation frequency
$\omega_{0}$ of the impurities too. The dependence is described by
an effectiveness function $\Xi\left(\omega_{0},
T_{\texttt{C}}^{\ast}\right)$. The function determines some
optimal frequency to obtain the critical temperature
$T_{\texttt{C}}^{\ast}$ with minimal concentration of the
impurities: $\omega_{0}=\pi T_{\texttt{C}}^{\ast}$. In limit cases
$T\gg\omega_{0}$ and $\omega_{0}\rightarrow\infty$ effectiveness
of the impurities aspires to zero, because at too small frequency
a thermal noise destroys the changes of oscillators' states by
electrons, and at too large frequency an interaction with the
impurities is weak. The increase of the critical temperature is a
result of a combined consistent action of metal's phonons and
impurities' oscillation upon electrons under the condition of
averaging over a disorder, where electrons move in Gauss random
field with a white noise correlator. The amplification of the
pairing takes place at presence of initially attractive
interaction between electrons in a matrix only. Thus the
impurities play a role of catalyst of superconductivity.

Estimation of $T_{\texttt{C}}^{\ast}$ shows that the critical
temperature can essentially exceed critical temperature of the
pure metal under the condition of optimal choice of parameters of
the matrix and the impurities. So for a matrix of $\texttt{Al}$ at
the parameter value $\frac{1}{k_{F}l}\approx 0.01\ll 1$ the
critical temperature reaches giant (room) values
$T_{\texttt{C}}^{\ast}\sim 300\texttt{K}$ in comparison with
critical temperature of the pure metal
$T_{\texttt{C}}=1.2\texttt{K}$ under the condition of optimal
relation between the critical temperature and the oscillation
frequency $\omega_{0}=\pi T_{\texttt{C}}^{\ast}$. Thus the
proposed model of the catalysis by impurities with retarded
interaction gives a principal possibility to obtain high critical
temperature at reasonable concentration of the impurities.

\appendix

\section{Elastic scattering by impurities} \label{A}

In a case of elastic scattering a impurity's potential is a
function of a wave vector only $\upsilon=\upsilon(\textbf{q})$. An
electron propagator is a function $G_{0}(\textbf{k},t_{2}-t_{1})$
and its Fourier-transform is
\begin{eqnarray}\label{A.1}
G_{0}(\textbf{k},\varepsilon)=\int_{-\infty}^{+\infty}d(t_{2}-t_{1})
e^{i\varepsilon(t_{2}-t_{1})}G_{0}(\textbf{k},t_{2}-t_{1})=\frac{1}{\varepsilon-\xi(k)+i\delta
\texttt{sign}\xi},
\end{eqnarray}
where $\xi(k)=\frac{k^{2}}{2m}-\varepsilon_{F}\approx
v_{F}(k-k_{F})$ is energy of an electron counted from Fermi
surface, $\varepsilon$ is an energy parameter, the series
(\ref{1.8}) can be represented in a form of Dyson equation:
\begin{equation}\label{A.2}
iG(\textbf{k},\varepsilon)=iG_{0}(\textbf{k},\varepsilon)+iG_{0}(\textbf{k},\varepsilon)(-i)\Sigma
iG(\textbf{k},\varepsilon),
\end{equation}
where  $\Sigma(\textbf{k},\varepsilon)$ is a mass operator. For a
weak disorder $\frac{1}{k_{F}l}\ll 1$ ($l$ is a free length) it is
possible to neglect the cross diagrams and to write the mass
operator in a form (Fig.\ref{Fig1A}):
\begin{figure}[ht]
\includegraphics[width=12cm]{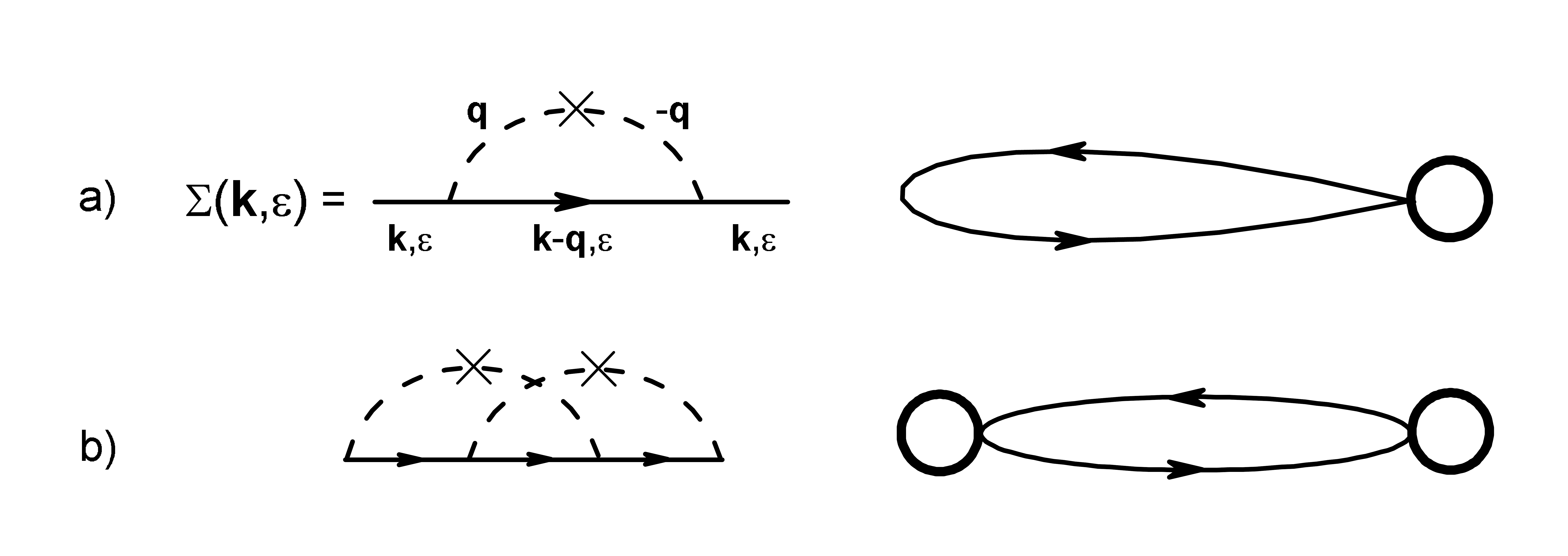}
\caption{Mass operators describing a multiple scattering of
electrons by impurities. Dotted lines with daggers on the diagrams
mean the scattering without an energy transfer, and a multiplier
$\rho\upsilon^{2}$ is related to them. A diagram (a) describes the
second Born approximation with an amplitude of the scattering
potential $\upsilon$. The diagram can be interpreted by the
picture on the right side - an infinite-to-one scattering by an
impurity. A cross diagram (b) describes a quantum correction to
the scattering - interference of incident and reflected by
impurities electron waves. This diagram can be interpreted as an
infinite-to-one scattering by two impurities with superposition of
the scattered waves.} \label{Fig1A}
\end{figure}
\begin{equation}\label{A.3}
-\Sigma(\textbf{k},\varepsilon_{n})=\rho\int\frac{d^{3}q}{(2\pi)^{3}}(-1)\upsilon(\textbf{q})
iG_{0}(\textbf{k}-\textbf{q},\varepsilon_{n})(-1)\upsilon(-\textbf{q})=\rho\int\frac{d^{3}p}{(2\pi)^{3}}|\upsilon(\textbf{k}-\textbf{p})|^{2}
iG_{0}(\textbf{p},\varepsilon_{n}),
\end{equation}
where we passed to Matsubara representation (where
$\varepsilon_{n}=(2n+1)\pi T$). It should be noted that in the
diagrams the dotted lines are not dressed with polarization loops,
because the disorder is "freezed in" and the impurities do not fit
into changes of an electron density. Substituting a free
propagator
$G_{0}(\textbf{p},\varepsilon_{n})=\frac{i}{i\varepsilon_{n}-\xi(p)}$
into the expression for a mass operator we obtaining (supposing a
weak dependence of the impurity's potential on momentum
$\upsilon(\textbf{k}-\textbf{p})\approx\upsilon$ and a linear
specter of quasi-particles near Fermi surface $\xi(k)\approx
v_{F}(k-k_{F})$):
\begin{eqnarray}\label{A.5}
\Sigma(\textbf{p},\varepsilon_{n})=-i\frac{\varepsilon_{n}}{|\varepsilon_{n}|}\pi\rho\upsilon^{2}\nu_{F}\equiv-i\gamma
\texttt{sign}\varepsilon_{n} \nonumber\\
G(\textbf{k},\varepsilon_{n})=\frac{1}{G_{0}^{-1}+i\Sigma}=\frac{i}{i\varepsilon_{n}-\xi(p)+i\gamma
\texttt{sign}\varepsilon_{n}}
\end{eqnarray}
where $\nu_{F}=\frac{mk_{F}}{2\pi^{2}}$ is a density of states on
Fermi surface per one projection of spin. Then the mean free time
and the free length are determined as \citep{sad}:
\begin{equation}\label{A6}
    \tau=\frac{1}{2\gamma},\qquad
    l=v_{F}\tau=\frac{v_{F}}{2\gamma}=\frac{v_{F}}{2\pi\rho\upsilon^{2}\nu_{F}}
\end{equation}
Elastic impurities do not influence upon effective mass of
quasi-particles but they condition a quasi-particles' damping
$\gamma\texttt{sign}\varepsilon_{n}$.

A small parameter for the perturbation theory is a ratio of a
contributions of cross diagrams to to a contribution of diagrams
without crossings \citep{sad}. Due scattering a momentum of an
electron obtains an uncertainty $\triangle k\sim 1/l$.  Then the
small parameter is $\frac{\triangle
k}{k_{F}}\sim\frac{1}{k_{F}l}$. Since in metals $1/k_{F}\simeq a$
(where $a$ is a lattice constant) then a weak disorder corresponds
to $\frac{1}{k_{F}l}\ll 1$. All diagrams with crossings describe
quantum corrections for conductivity - interference of incident
and reflected by impurities electron waves. This results to
Anderson's localization - transition of a metal to an insulator
state \citep{sad,sad1,pat} when $\frac{1}{k_{F}l}\gtrsim 1$
(electrons are "blocked" between the impurities). However with
increase of temperature (or if the system is in an external
alternating field \citep{altsh1}) processes of a nonelastic
scattering begin to play a role (electron-phonon processes -
\citep{altsh2}, electron-electron processes - \citep{altsh3}). The
processes limit the coherence time of electron waves
$\tau_{\varphi}<\infty$ (or the coherence length
$L_{\varphi}<\infty$). If $\tau_{\varphi}<\tau$ (or
$L_{\varphi}<l$) then the interference contribution is essentially
suppressed (the phase failure takes place) \citep{levit}, and the
cross diagram can be neglected.

\end{document}